\documentclass{PoS}
\usepackage{lineno}
\usepackage{amsmath}
\newcommand{\sqrtsNN}{\sqrt{s_{\rm \scriptscriptstyle NN}}}

\newcommand{\GeV}{\mathrm{GeV}}
\newcommand{\TeV}{\mathrm{TeV}}

\newcommand{\PbPb}{\mbox{Pb--Pb}}

\newcommand{\Raa}{R_{\rm AA}}
\newcommand{\Taa}{T_{\rm AA}}
\newcommand{\pt}{p_{\rm T}}
\newcommand{\ptch}{p_\text{T, ch jet}}
\newcommand{\ptchraw}{p_\text{T, ch jet}^{\rm raw}}
\newcommand{\ptchrawkt}{p_\text{T, ch jet}^{{\rm raw}, k_{\rm T}}}
\newcommand{\DtoKpi}{{\rm D}^0 \to {\rm K}^-\pi^+}
\newcommand{\DtoKpipi}{{\rm D}^+\to {\rm K}^-\pi^+\pi^+}

\newcommand{\DstartoDpi}{{\rm D}^{*+} \to {\rm D}^0 \pi^+ }

\newcommand{\Dzero}{\rm D^0}

\newcommand{\Dstar}{{\rm D^{*+}}}

\newcommand{\Dplus}{{\rm D^+}}

\newcommand{\Jpsi}{{\rm J}/\psi}

\newcommand{\de}{{\rm d}}
\newcommand{\mum}{\mu{\rm m}}
\newcommand{\qtwoTPC}{q_2^{\rm TPC}}
\title{Study of $\pmb{\Raa}$ and $\pmb{v_2}$ of non-strange D mesons and D-jet production in Pb--Pb collisions with ALICE}

\ShortTitle{$\Raa$ and $v_2$ of D mesons and D-jet production in Pb--Pb collisions with ALICE}

\author{\speaker{Fabrizio Grosa} on behalf of the ALICE Collaboration\\
        Politecnico di Torino, Corso Duca degli Abruzzi 24, 10129 Torino Italy\\
        INFN sez. Torino, via Pietro Giuria 1, 10125 Torino Italy\\
        E-mail: \email{fabrizio.grosa@cern.ch}}

\abstract{Heavy quarks are sensitive probes of the colour-deconfined medium formed in ultra-relativistic heavy-ion collisions, the Quark--Gluon Plasma (QGP). The ALICE Collaboration measured the production of $\Dzero$, $\Dplus$, and $\Dstar$ mesons in $\PbPb$ collisions at $\sqrtsNN=5.02~\TeV$. The properties of the in-medium energy loss are investigated via the measurement of the nuclear modification factor ($\Raa$) of non-strange D mesons. The modification of the D-meson  transverse momentum ($\pt$) distributions inside the jet is studied via the measurement of the D-meson tagged jet $\Raa$ in central $\PbPb$ collisions. In mid-central collisions, the measurement of the D-meson elliptic flow ($v_2$) at low and intermediate $\pt$ gives insight into the participation of the charm quark in the collective motion of the system, while at high $\pt$ it constrains the path-length dependence of the energy loss. The coupling of the charm quark to the light quarks in the underlying medium is further investigated with the application of the event-shape engineering (ESE) technique to D-meson elliptic flow.}

\FullConference{
European Physical Society Conference on High Energy Physics - EPS-HEP2019 -\\
			10-17 July, 2019\\
			Ghent, Belgium}

\begin{document}

\section{Introduction}
In ultra-relativistic heavy-ion collisions, heavy quarks (charm and beauty) are produced in the early times of the reaction via hard-scattering processes and they subsequently probe the colour-deconfined medium, known as Quark--Gluon Plasma (QGP). The measurement of the nuclear modification factor ($\Raa$) of hadrons containing heavy quarks, defined as the ratio between the $\pt$-differential yields in nucleus--nucleus collisions ($\de N_{\rm AA}/\de\pt$) and the $\pt$-differential production cross section measured in pp collisions ($\de \sigma_{\rm pp}/\de\pt$) scaled by the average nuclear overlap function $\langle\Taa\rangle$, is used to study the properties of the in-medium parton energy loss. The comparison to light-flavour hadrons provides information about the quark-mass and colour-charge dependence, while the measurement of the charm jets can be used to study the modification of the internal jet sub-structure in the medium.

Further insights into the interaction of heavy quarks with the QGP are given by the measurement of azimuthal anisotropies, which are typically characterised in terms of Fourier coefficients, $v_{\rm n}=\langle \cos{\rm n}(\varphi-\Psi_{\rm n})\rangle$, where $\varphi$ is the particle-momentum azimuthal angle, the brackets denote the average over all the measured particles in the considered events, and $\Psi_{\rm n}$ is the symmetry-plane angle relative to the n$^{th}$ harmonic. The second-harmonic coefficient, called elliptic flow, is the dominant term in mid-central heavy-ion collisions due to the almond-shaped interaction region. At low $\pt$ it is sensitive to the participation in the collective dynamics of the underlying medium and the degree of thermalisation of the heavy quarks in the medium, while at high $\pt$ it is governed by the path-length dependence of the parton energy loss in the medium.

\section{D-meson and jet reconstruction}
Open-charm production in Pb--Pb collisions at $\sqrtsNN=5.02~\TeV$ was measured by ALICE via the exclusive reconstruction of D mesons at mid-rapidity ($|y|<0.8$), in the hadronic decay channels $\DtoKpi$ ($c\tau\simeq 123~\mum$, ${\rm BR}=3.89\%$), $\DtoKpipi$ ($c\tau\simeq 312~\mum$, ${\rm BR}=8.98\%$), and $\DstartoDpi\rightarrow {\rm K^-\pi^+\pi^+}$ (strong decay, ${\rm BR}=2.63\%$)~\cite{Tanabashi:2018oca}. The decay topologies were reconstructed exploiting the excellent vertex-reconstruction capabilities of the Inner Tracking System (ITS). Kaons and pions were identified with the Time Projection Chamber (TPC) via their specific energy loss and with the Time-Of-Flight detector (TOF). The raw D-meson yields were extracted via an invariant-mass analysis after having applied topological selections to enhance the signal over background ratio. The efficiency-times-acceptance corrections were obtained from MC simulations based on \textsc{hijing}~\cite{Wang:1991hta} and \textsc{pythia6}~\cite{Sjostrand:2006za} event generators and the \textsc{geant3} transport package~\cite{Brun:1994aa}. The fraction of prompt D mesons was estimated with a \textsc{fonll}-based approach \cite{Cacciari:1998it,Acharya:2017qps}. The centrality and the direction of the event plane (estimator of $\Psi_2$) were provided by the V0 scintillators, which cover the pseudorapidity regions $-3.7 < \eta < -1.7$ (V0C) and $2.8 < \eta < 5.1$ (V0A). The measurement of the D-meson $v_2$ was performed with the scalar-product (SP) method~\cite{Voloshin:2008dg}. The $\Dzero$-meson tagged charged jets were reconstructed with the \textsc{fastjet} package~\cite{Cacciari:2011ma} and the anti-$k_{\rm T}$ clustering algorithm~\cite{Cacciari:2008gp} using charged-particle tracks and requiring the presence of the daughter tracks of a $\Dzero$ meson among the jet constituents. The covered charged-jet transverse-momentum range was $5<\ptch<20~\GeV/c$, with the jet resolution parameter $R=0.3$ and a requirement of the D-meson transverse momentum $p_\text{T,D}>3~\GeV/c$. The background density scaled by the area of the reconstructed signal jet was subtracted from the reconstructed $\pt$ of the signal jet. The underlying background momentum density was estimated event-by-event using the median of $\ptchrawkt/A_{\rm jet}$, where $\ptchrawkt$ is the uncorrected $\ptch$ and $A_{\rm jet}$ is the area of the jets reconstructed with the $k_{\rm T}$ algorithm. The $\ptchraw$ spectrum of the jets reconstructed with the anti-$k_{\rm T}$ algorithm was then unfolded for the detector response and the background fluctuations in the underlying event extracted using the random cone method~\cite{Abelev:2012ej}.

\section{D-meson nuclear modification factor}
\begin{figure}[tb]
\centering
\includegraphics[width=0.48\textwidth]{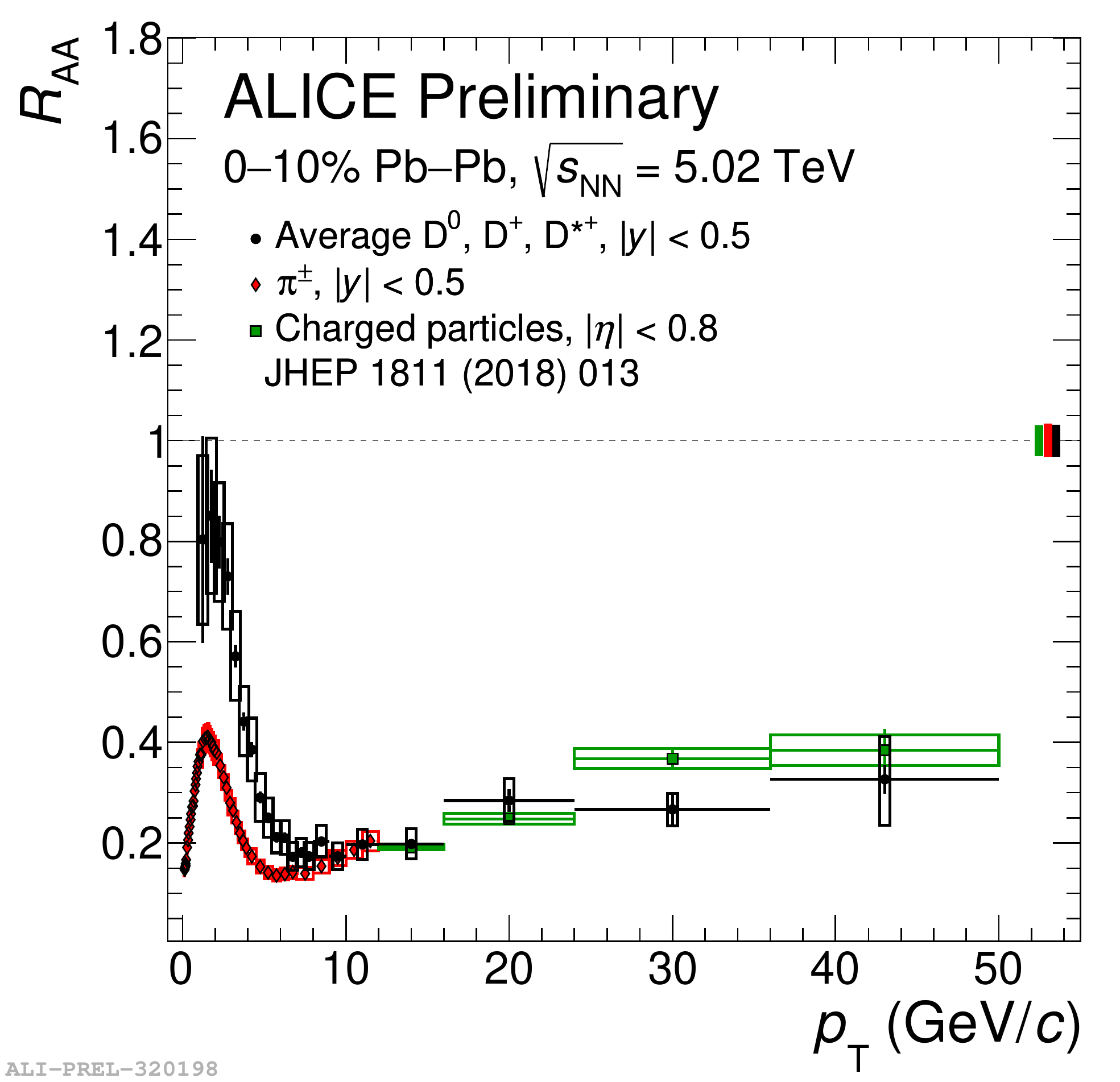}
\includegraphics[width=0.48\textwidth]{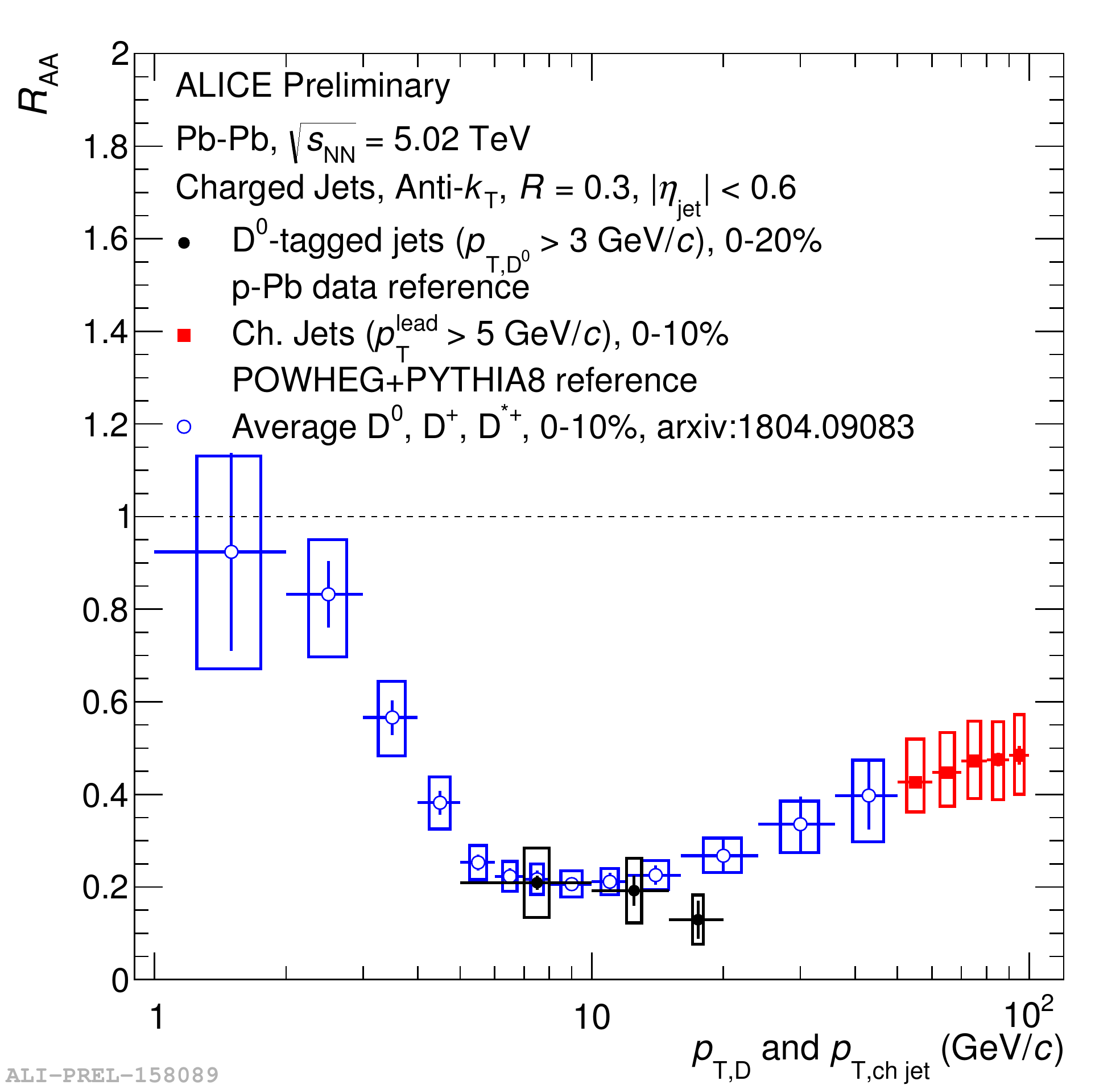}
\caption{Average prompt $\Dzero$, $\Dplus$, and $\Dstar$ $\Raa$ in Pb--Pb collisions at $\sqrtsNN=5.02~\TeV$ in the 0--10\% centrality class compared to the $\Raa$ of $\pi^\pm$ for $\pt<12~\GeV/c$ and charged particle $\Raa$ for $\pt>12~\GeV/c$ (left panel) and to the $\Raa$ of charged jets and $\Dzero$-meson tagged jets in the 0--20\% centrality class (right panel).}
\label{fig:DmesonRaa}      
\end{figure}

The left panel of Fig.~\ref{fig:DmesonRaa} shows the average $\Raa$ of prompt $\Dzero$, $\Dplus$, and $\Dstar$ mesons measured in the 10\% most central Pb--Pb collisions at $\sqrtsNN=5.02~\TeV$. It is compared to the charged-pion and charged-particle~\cite{Acharya:2018qsh} $\Raa$ measured at the same energy and centrality class for $\pt<12~\GeV/c$ and $\pt>12~\GeV/c$, respectively. The $\Raa$ of D mesons is higher than that of charged pions by more than $2\,\sigma$ in each $\pt$ bin for $\pt<6~\GeV/c$, suggesting a quark-mass dependence of the in-medium energy loss. This observation is however nontrivial, since several other effects can contribute to this difference, such as the different scaling of soft and hard probes at low $\pt$, the different initial shapes of $\pt$ spectra, different contributions of radial flow and hadronisation via recombination.   

In the right panel of Fig.~\ref{fig:DmesonRaa}, the $\Raa$ of $\Dzero$-tagged jets is shown. A strong suppression, similar to that of the prompt D-meson~\cite{Acharya:2018hre} is observed.

\begin{figure}[tb]
\centering
\includegraphics[width=0.48\textwidth]{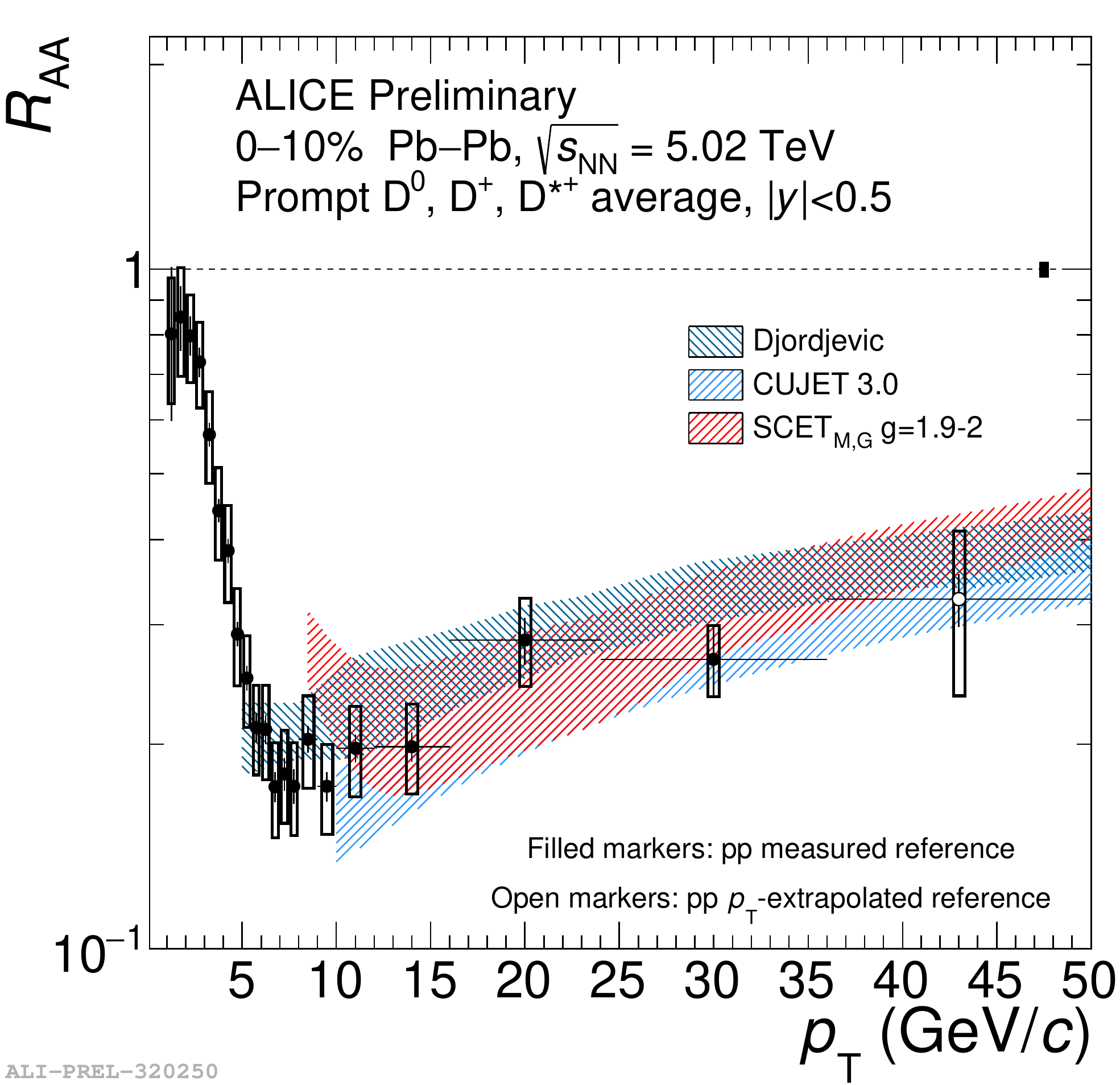}
\includegraphics[width=0.48\textwidth]{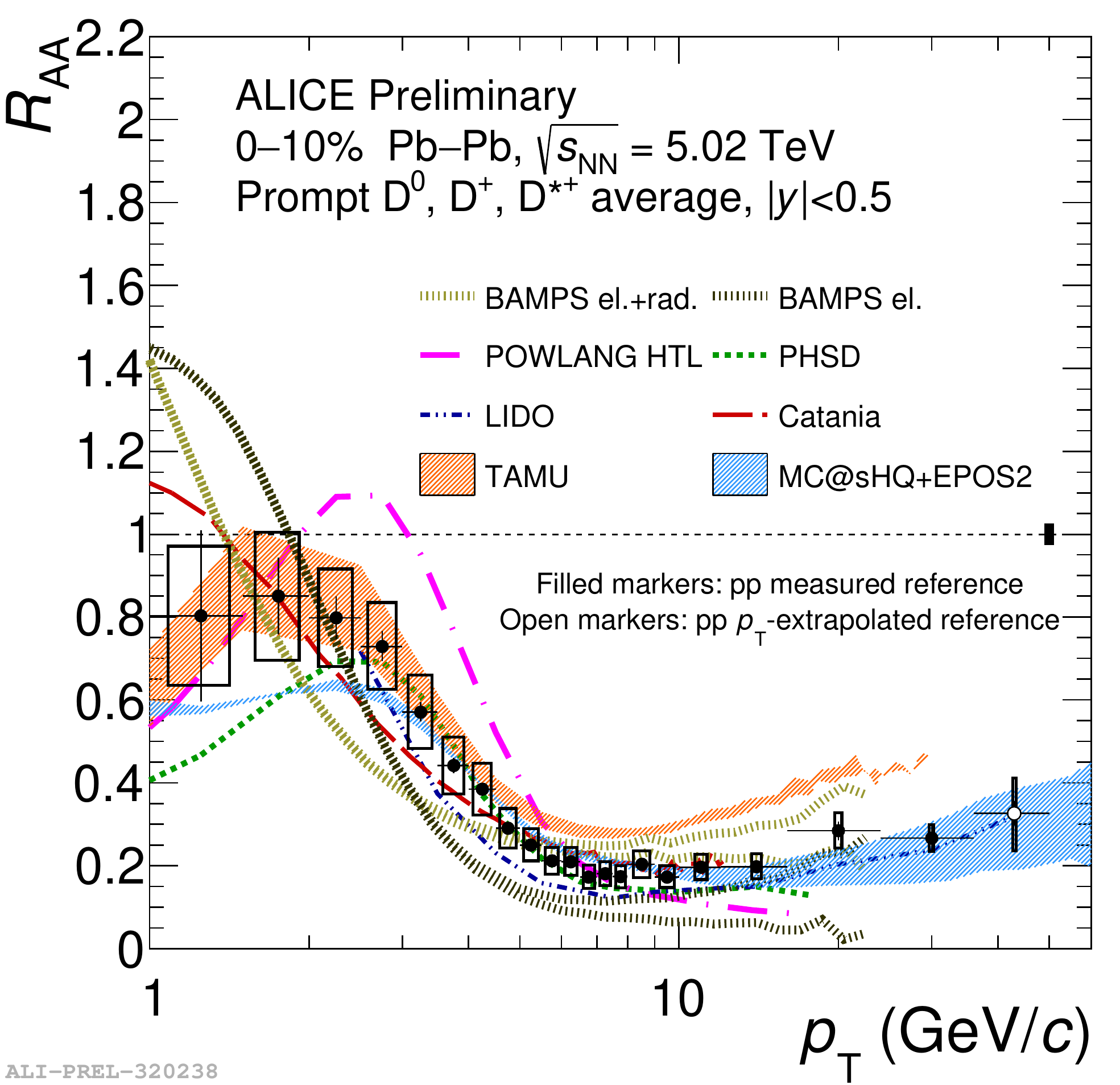}
\caption{Average prompt $\Dzero$, $\Dplus$, and $\Dstar$ $\Raa$  in Pb--Pb collisions at $\sqrtsNN=5.02~\TeV$ in the 0--10\% centrality class compared to model predictions based on pQCD calculations of parton energy loss (left panel) and on the charm-quark transport in an expanding medium (right panel).}
\label{fig:DmesonRaaVsModels}      
\end{figure}

Figure~\ref{fig:DmesonRaaVsModels} shows the comparison between the measured D-meson $\Raa$ and that predicted by theoretical models based on perturbative QCD (pQCD) calculations of high $\pt$ charm-quark energy loss~\cite{Djordjevic:2013pba, Xu:2015bbz, Kang:2016ofv} (left panel) and charm-quark transport in a hydrodynamically expanding medium~\cite{Aichelin:2012ww, Song:2015ykw, Ke:2018tsh, Uphoff:2014hza, He:2014cla, Beraudo:2014boa} (right panel). The $\Raa$ for $\pt>10~\GeV/c$ is well described by pQCD-based models, as well as by the \textsc{mc}@s\textsc{hq+epos2} and \textsc{lido} models. The low $\pt$ region of the measured $\Raa$ is fairly reproduced by most of the models that implement charm-quark transport.
 
\section{D-meson azimuthal anisotropies}
\begin{figure}[tb]
\centering
\includegraphics[width=0.65\textwidth]{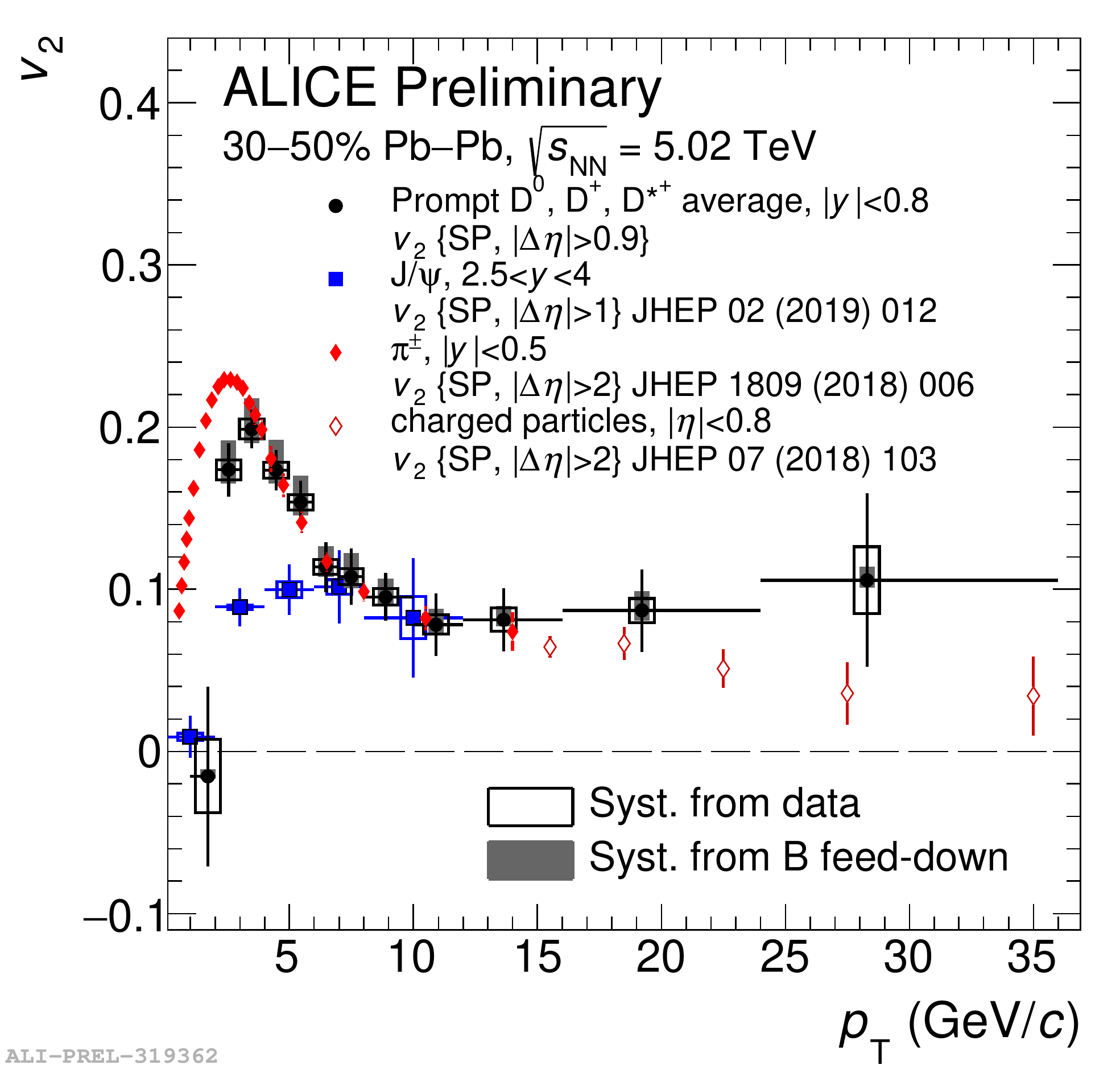}
\caption{Average prompt $\Dzero$, $\Dplus$, and $\Dstar$ $v_2$ in Pb--Pb collisions at $\sqrtsNN=5.02~\TeV$ in the 30--50\% centrality class, compared to that of charged pions~\cite{Acharya:2018zuq} for $\pt<15~\GeV/c$, charged particles~\cite{Acharya:2018lmh} for $\pt>15~\GeV/c$ at mid-rapidity, and inclusive $\Jpsi$ mesons at forward rapidity~\cite{Acharya:2018pjd}.}
\label{fig:Dmesonv2}      
\end{figure}
Figure~\ref{fig:Dmesonv2} shows the average $v_2$ of prompt $\Dzero$, $\Dplus$, and $\Dstar$ measured in mid-central (30--50\%) $\PbPb$ collisions at $\sqrtsNN=5.02~\TeV$, compared to that of charged pions~\cite{Acharya:2018zuq} for $\pt<15~\GeV/c$, charged particles~\cite{Acharya:2018lmh} for $\pt>15~\GeV/c$ at mid-rapidity, and inclusive $\Jpsi$ mesons at forward rapidity~\cite{Acharya:2018pjd}.
 
The average prompt D-meson $v_2$ was found to be lower (similar) to that of charged pions for $\pt<3~\GeV/c$ ($\pt>3~\GeV/c$) and higher than that of $\Jpsi$ mesons for $\pt<6~\GeV/c$. This observation is consistent with the scaling of the $v_2$ with the mass of the meson species observed for light-flavour hadrons below $3~\GeV/c$~\cite{Acharya:2018zuq}, and the increase of the D-meson $v_2$ due to the hadronisation of the charm quark via recombination with flowing light-flavour quarks~\cite{Nahrgang:2014vza}. The measured $v_2$ coefficients for all the meson species converge to a similar value for $\pt>6-8~\GeV/c$, as expected when the path-length dependence of the in-medium parton energy loss is the mechanisms that originates the azimuthal anisotropy.

The $v_2$ of $\Dzero$, $\Dplus$, and $\Dstar$ mesons was further investigated applying the event-shape engineering (ESE) technique~\cite{Voloshin:2008dg}. This technique relies on the classification of events with fixed centrality but different average elliptic flow, quantified by the magnitude of the second-harmonic reduced flow vector, $\qtwoTPC=|\pmb{Q}_2^{\rm TPC}|/\sqrt{M}$, where $M$ is the multiplicity and $\pmb{Q}_2^{\rm TPC}$ is the second-harmonic flow vector measured with charged-particle tracks reconstructed in the TPC detector having $|\eta|<0.8$~\cite{Acharya:2018bxo}.

The D-meson $v_2$ was found to be higher (lower) in the 20\% of events with largest (smallest) $\qtwoTPC$, confirming a correlation between the D-meson azimuthal anisotropy and the collective expansion of the bulk matter. 

\begin{figure}[tb]
\centering
\includegraphics[width=1.\textwidth]{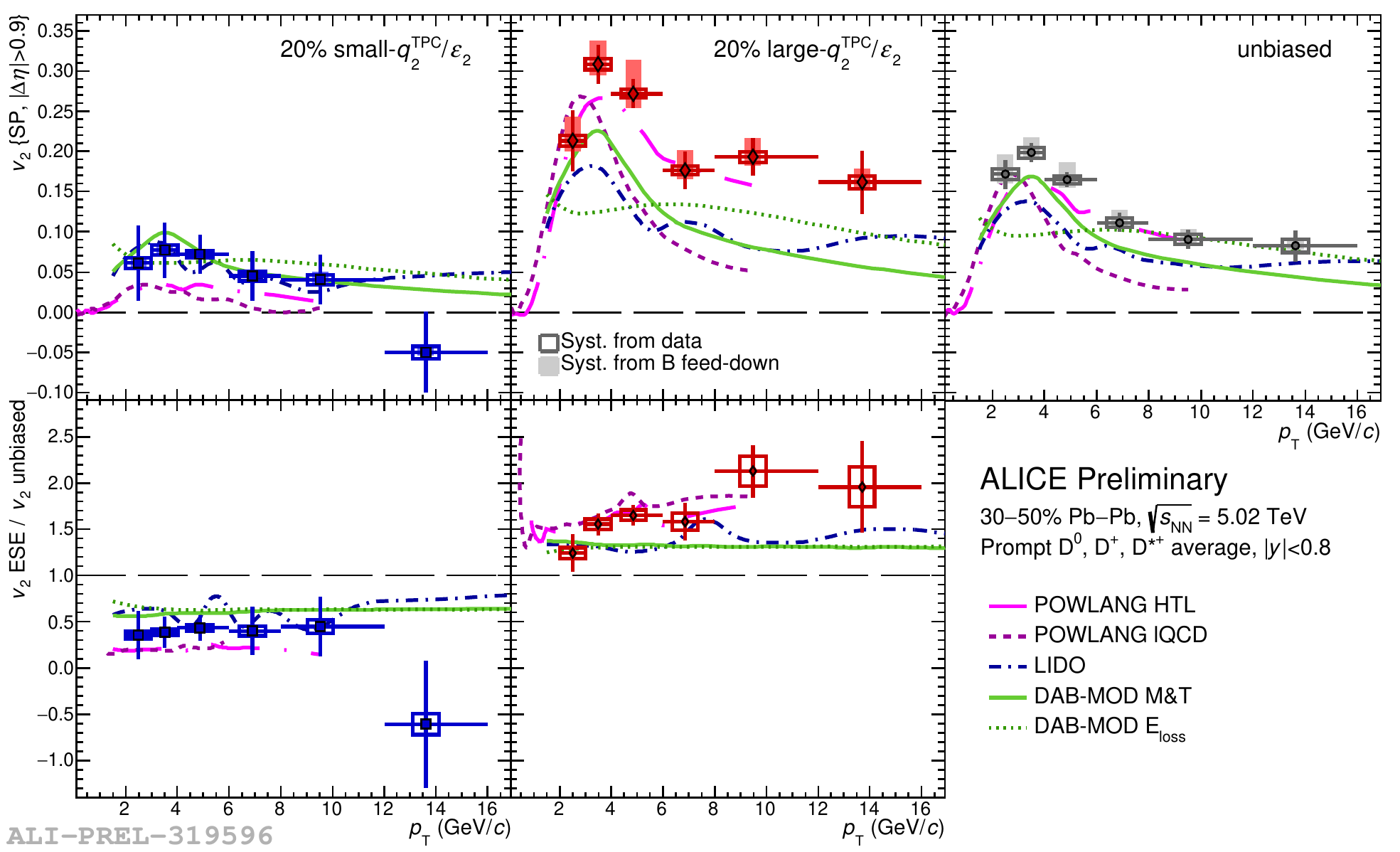}
\caption{Average prompt $\Dzero$, $\Dplus$, and $\Dstar$ $v_2$ in Pb--Pb collisions at $\sqrtsNN=5.02~\TeV$ in the 30--50\% centrality class, for the small-$\qtwoTPC$ (top-left panel), large-$\qtwoTPC$ (top-middle panel), and unbiased (top-right panel) samples compared to transport models. The ratios between the $v_2$ in the ESE-selected and unbiased samples are displayed in the bottom panels.}
\label{fig:Dmesonv2ESEVsModels}      
\end{figure}

The top panels of Fig.~\ref{fig:Dmesonv2ESEVsModels} show the average $v_2$ of prompt D mesons measured in the samples of events with 20\% smallest $\qtwoTPC$, 20\% largest $\qtwoTPC$, and without ESE selection (unbiased sample) compared to the predictions provided by models that implement the charm-quark transport in a hydrodynamically expanding medium. In the bottom panels of the same figure, the ratios of the $v_2$ in the ESE-selected and unbiased samples are displayed. The \textsc{powlang}~\cite{Beraudo:2018tpr} model describes the measurements in the unbiased and 20\% large-$\qtwoTPC$ samples, while it underestimates the measurement in the small-$\qtwoTPC$ sample. Conversely, the \textsc{dab-mod}~\cite{Katz:2019fkc} and \textsc{lido}~\cite{Ke:2018tsh} models better describe the measured D-meson $v_2$ in the small-$\qtwoTPC$ sample and underestimate those in the unbiased and large-$\qtwoTPC$ samples. 
  
\section{Conclusions}
In this contribution, the most recent results on the production and azimuthal anisotropy of D mesons, measured in $\PbPb$ collisions at $\sqrtsNN=5.02~\TeV$, were presented. The measurements of the D-meson $\Raa$ and $v_2$, performed on the latest sample of Pb--Pb collisions collected in 2018, have an improved statistical precision by about a factor three (two) in the 0--10\% (30--50\%) centrality class compared to the previous results published by the ALICE Collaboration~\cite{Acharya:2017qps, Acharya:2018hre, Acharya:2018bxo}. The first measurement of the $\Dzero$-meson tagged jet $\Raa$ in Pb--Pb collisions, performed with the data sample collected in 2015, was also presented. The precision of the measurement and the $\ptch$ reach will be greatly improved with the sample of Pb--Pb collisions collected in 2018.

\bibliographystyle{utphysmy} 
\bibliography{DmesonPbPb_EPS_Grosa}

\end{document}